\documentclass[11pt, reqno]{amsart}

\usepackage{amsmath,amsfonts}
\usepackage{latexsym,amssymb}
\usepackage{bm}
\usepackage[dvipsnames]{xcolor}

\begin{document}

\title[]{Comment on work of Umrigara and Anderson ``Energy needed to propel a tiny spacecraft to Proxima Centauri, and, an unstated assumption in Einstein's 1905 paper'' (arXiv:2502.04331v1)}

\author{V.~Onoochin}

\date{}

\begin{abstract}

In a recent paper by Umrigaral and Anderson, the authors make an important statement: it is possible to misunderstand the behavior of physical objects (parts of a spacecraft) if that object has reached a velocity comparable to the speed of light. Achieving such a velocity for a spacecraft is critical for interstellar travel, since only traveling at such speeds makes travel to distant stars possible within a human lifetime.

Last years, several ideas have been proposed to achieve relativistic velocities for a spacecraft. The authors of these ideas understand that this task is very difficult due to technological obstacles. However, when describing these projects, the authors omit mentioning possible obstacles of physical origin. The paper by Umrigaral and Anderson is the first to consider physical limitations on construction of future spacecraft.

In this Comment, it will be shown that, according to Umrigaral and Anderson, not only technological obstacles will arise in the project of interstellar flights, but also the physical limitations of a spacecraft moving at relativistic velocities may become more serious problems for the implementation of such a project. 

\end{abstract}

\maketitle

Although human progress in exploring the Solar system is very modest, there are some projects to develop a spacecraft that could make interstellar flights within a human lifetime. Usually, the authors of such ideas understand the obstacles to the implementation of these projects. However, it is generally accepted that these obstacles are considered as of technological type and will be overcome due to further progress. Therefore, the significance of the work of Umrigaral and Anderson~\cite{UA} is that they point out the presence of obstacles of physical origin.

As the authors correctly estimate, the only option for achieving such velocities for a spacecraft is to use a sail or some kind of adjustable mirror that should receive a light beam from a remote powerful laser, reflect this beam and obtain some mechanical momentum due to this reflection.

However, despite the simplicity of this technical solution (it is assumed that a sufficiently powerful beam will be supplied by a laser), some obstacles to this solution are present. For example, the frequency of the reflected photons changes due to the Doppler effect when the spacecraft and its mirror accelerate. This can reduce the magnitude of the resulting mechanical moment.

However, the authors neglect to consider the impact of other physical factors that can hinder the acceleration of spacecraft to relativistic velocities. Let us consider these factors.

Umrigaral and Anderson assume that a {\it perfect mirror} will be mounted in the spacecraft. But this assumption means that physical parameters of the spacecraft reflector remain constant as the velocity of the apparatus increases. It is essential to assess the impact of velocity increase on these parameters.

Reflection layer of the mirror is made of a material with high electrical conductivity, which ensures a perfect (close to 1) coefficient of reflection. The higher the electrical conductivity, $\sigma$, the better the incident EM wave induces the electric current on the metal surface of the mirror. This induced electric current produces a magnetic field that compensates the magnetic field of the incident wave, as shown in~\cite{Symp}. This ensures that the incident wave does not penetrate the metal; instead, the reflected EM wave is created by this induced current.

This mechanism works perfectly for high value of the conductivity. Hence, mirrors made by means metals like silver have the reflection coefficient close to 1. The conductivity in prefect metals is described by the expression~(Eq.~(7.58) of~\cite{JDJ}),
\begin{equation}
\sigma=\frac{f_0Ne^2}{m(\gamma_0 -i\omega)}\,,
\end{equation}
where $m$ is the mass of electron, $f_0N$ the number of free electrons per unit volume in the medium. The damping constant $\gamma_0/f_0$ is determined empirically from experimental data on the conductivity. For a mirror moving with relativistic velocities all parameters of this formula are the same as for the mirror at rest. But the mass of the electrons increases while their velocity increases as 
\begin{equation}
m=\frac{m_0}{\sqrt{1-(v/c)^2}} \,.
\end{equation}
To preserve historical justness, let us note that this formula was introduced by Lorentz~\cite{Lor1904}) . So at $v=0.5c$ the increase of the electron masses give to decrease of the conductivity $\sigma_{0.5c}=0.85\sigma_0$. It decreases the reflection coefficient.

It can be argued that in the frame of reference co-moving the spacecraft, the mass of electrons should be the same as that of electrons at rest. But there are two counterarguments to this statement.\newline
$\square$ the magnitude of the transferred mechanical momentum is calculated as the difference in the momenta of the incident and reflected beams. At least, the parameters of the incident beam are determined in the inertial system of the Earth (or the Solar system, where the laser is located);\newline
$\square$ while the laser beam reaches the mirror, the spacecraft accelerates. Therefore, the co-moving frame is not inertial, and no assumptions can be made about the masses of electrons in this system.

Meanwhile, when the apparatus acquires relativistic velocities, not only the reflection coefficient decreases. Any macroscopic body moving with some 'constant' velocity~\footnote{The velocity $V$ of some body can be considered constant if the product of the acceleration of this body $a$ and the duration of the considered time $t$ is much less than the velocity, $at\ll V$. } can be considered as a set of particles (electrons and ions forming the structure of the body) moving in parallel courses. Then, as noted in Chapter 19-4~\cite{PP}, the Coulomb potential acting between the charged particles changes to the convection potential $\Psi$,
\begin{equation}
\Psi = \frac{1-v^2/c^2}
{\sqrt{(x-x'^2)+(1-v^2/c^2)\left[(y-y')^2+(z-z')^2\right]}} \,\,,
\label{Psi}
\end{equation}
where $x',\,y',\,z'$ are coordinates of the moving charge.

Since the electrostatic potential determines the equilibrium points of ions in the crystal lattice of a certain solid when the latter is at rest, it can be assumed that if this body is in motion, then the convective potential should determine the equilibrium points of ions also in this case. It follows from Eq.~(\ref{Psi}) that when the velocity of the body changes, the value of the convection potential also changes. Therefore, the equilibrium positions of the ions must also change. Let us find out how the new equilibrium points depend on the body velocity. 

To do this, it would be reasonable to consider the motion of some solid body consisting of an ionic crystal (for example, a NaCl crystal). This choice of body material is due to the intention to reduce the analysis of the behavior of a solid body to the analysis of electrostatic forces that ensure equilibrium of the crystal lattice. Indeed, it is possible to calculate the change in the equilibrium points of the lattice ions of some perfect metal. In this case, the distribution of conduction electrons in the lattice should be taken into account. The dominant factor determining the location of ions at lattice sites is the force of electrostatic repulsion between ions, but this force is shielded by a negative spatial charge due to conduction electrons. So even in this case the problem can be reduced to an electrostatic one. This can be established from considering the Hamiltonian of the crystal (Chapter~1.3 of the book~\cite{Z}, Eq. (1.3.1))  
\begin{equation}
H=\frac{1}{2}\sum \limits_l\frac{\bm{p}_l^2}{m}+
U(\bm{R}_1,\bm{R}_2,...\bm{R}_l,...)\,\,,
\end{equation}
where the potential energy of the crystal is only a function of distances $\bm{R}_1,\bm{R}_2,...\bm{R}_l,...$ between the atoms of the lattice  (Eq.~(1.3.3) of~\cite{Z}). Thus, in the equilibrium configuration, when the atoms are located exactly at lattice sites, one obtains  
\begin{equation}
\frac{\partial U}{\partial \bm{r}_l}=0\quad;\quad
\bm{r}_1=\bm{r}_2=...=\bm{r}_N=0  \label{0}
\end{equation}
for all $\bm{r}_l=\bm{R}_l-l\bm{a}$, $\bm{a}$ is the vector of elementary lattice cell. In the solid state problems, when
studying the dynamical properties of a crystal, the potential energy is represented as  
\begin{equation}
U(\bm{r}_1,\bm{r}_2,...\bm{r}_l,...)=U_0+\sum \limits_{l,l'}\bm{r}_l
\bm{r}_{l'}\frac{\partial^2 U}{\partial \bm{r}_l\partial \bm{r}_{l'}}
\,\,,
\end{equation}
where $U_0$ is the minimum of the electrostatic energy of NaCl crystal. Let us assume that when the lattice is being at rest, the distance between two neighbor ions is $d$.  Then $U_0$ for the cubic lattice,
\begin{equation}
U_0=W_{rest}=-\sum \limits_{\substack{l,\,m,\,n=0\\l+m+n\neq0}}^{\infty} \frac{(-1)^{l+m+n}2^3e^2}{\sqrt{(ld)^2+(md)^2+(nd)^2}}=
-\frac{2^3e^2}{d}\sum \limits_{\substack{l,\,m,\,n=0\\l+m+n\neq0}}^{\infty} \frac{(-1)^{l+m+n}}{\sqrt{l^2+m^2+n^2}}\,\, , \label{W_r}
\end{equation}
where summation over $l$ corresponds to summation of the ions along the $x$ axis, summation over $m$ does along the $y$ axis and summation over $n$ does along the $z$ axis; the singular term with $l = m = n = 0$ is excluded. For the body at rest, only Coulomb potential acts between the ions of the lattice. 

When the crystal acquires some velocity, the Coulomb potential changes to the convection potential. The forces, which are calculated as $-\nabla \Psi$ and which act between ions, change and the distances between ions must change in order to ensure a minimum of the potential energy of the lattice. It is reasonable to assume that since the process of crystal acceleration is adiabatic (much slower than the ability of ions to move) this process conserves constancy of the minimum of the potential energy. 

If the Coulomb potential in Eq.~\eqref{W_r} changes to the convention potential, Eq.~\eqref{Psi}, the potential energy of the lattice becomes,
\begin{equation}
	W_{mov}=-\sum \limits_{\substack{l,\,m,\,n=0\\l+m+n\neq0}}^{\infty}\frac{(-1)^{l+m+n}(1-v^2/c^2)2^3e^2}
	{\sqrt{(ld)^2+(1-v^2/c^2)\left[(md)^2+(nd)^2\right]}}\,\,,\label{W_m}
\end{equation}
Comparing Eq.~(\ref{W_r}) to Eq.~(\ref{W_m}) shows that the total internal energy of the lattice increases with the velocity. It means that the ions are not located at initial points of equilibrium and it is necessary to find new points of location of the ions in such a way that the magnitude of $W_{mov}$ will be equal to the value of $W_{rest}$. 

In ~\cite{O}, new ionic equilibrium points are determined by computer calculations; they correspond to changes in interatomic distances as
\begin{equation}
	d_{mov}^{\,\|}=(1-v^2/c^2)d_{rest}^{\,\|} \quad;
	\quad d_{mov}^{\bot}=\sqrt{1-v^2/c^2}d_{rest}^{\bot}  \,\,, \label{d_rel}
\end{equation}
This change of interatomic distances gives the change of dimensions of the whole moving body. In fact, it is like a so-called Lorentz contraction, but derived without involving any hypothesis. Also, this contraction is stronger than the relativistic contraction. But Eq.~\eqref{d_rel} are obtained by means of reliably verified principles of solid state physics and classical electrodynamics.

The above results predict that the spacecraft components will deform (change their dimensions) at relativistic velocities. This may not be so critical for mechanical parts. But it may be critical for the electronics of a spacecraft accelerated to speeds of $v\simeq 0.1c$. The basis of modern electronic devices are semiconductor structures, which are very sensitive to the presence of any defects. The deformation analyzed above should be considered as a strong defect, since it can damage the crystal lattice of the semiconductor material. This factor and the increase in the mass of carriers will lead to malfunctions in electronic devices.

Actually, this conclusion is made only on consideration what should change with material of the body moving with relativistic velocities. Only experimental verification of the predicted problems with functioning the working systems of the spacecraft can show how are real these problems. But from the other side, the existence of these problems should follow from the solid state physics and electrodynamics. If technological obstacles in realization of the interstellar flights can be overcome by the progress in the technology, the obstacles of the physical origin, such as increase of the electron masses and change of the interatomic distances of the material of the moving body, cannot be overcome.

Actually, this conclusion is drawn only on the basis of consideration of what must change with the material of a body moving at relativistic speeds. Only experimental verification of the predicted problems with the functioning of the working systems of the spacecraft can show how real they are. But on the other hand, the existence of these problems must follow from solid state physics and electrodynamics. While technological obstacles to interstellar flight can be overcome by advances in technology, obstacles of physical origin, such as increasing masses of electrons and changing interatomic distances of the material of a moving body, cannot be overcome.

\end{document}